\documentclass[useAMS]{mn2e}
\usepackage{graphicx,subfigure}
\usepackage{pslatex}

\title[Vertical stratification of iron in atmospheres of BHB stars]{Vertical stratification of iron in atmospheres of blue horizontal-branch stars}
\author[V. Khalack et al.]
       {V. Khalack,$^1$ F. LeBlanc,$^1$  B. B. Behr$^2$ \\
               $^1$D\'epartement de Physique et d'Astronomie,
               Universit\'e de Moncton, Moncton, N.-B., Canada E1A 3E9\\
               $^2$Department of Systems Design Engineering,
               University of Waterloo, Waterloo ON N2L 3G1, Canada
            }
\date{Accepted ???.
      Received ???;
      in original form ???}

\pagerange{\pageref{firstpage}--\pageref{lastpage}}
\pubyear{2010}

\begin{document}

\maketitle

\label{firstpage}

\begin{abstract}
The aim of this study is to search for observational evidence of vertical iron
stratification in the atmosphere of fourteen blue horizontal-branch (BHB) stars.
We have found from our numerical simulations that five BHB stars:
B22, B186 in the globular cluster NGC~288, WF2-820, WF2-2692 in M13 and
B203 in M15 show clear signatures of the vertical stratification of iron
whose abundance increases toward the lower atmosphere. Two other BHB stars (B334 in M15 and B176 in M92)
also show possible iron stratification in their atmosphere.
A dependence of the slope of iron stratification on the effective temperature was also discovered.
It is found that the vertical stratification of iron is strongest in BHB stars with $T_{\rm eff}$ around 11,500K.
The slope of iron abundance decreases as $T_{\rm eff}$ increases and becomes negligible
for the BHB stars with $T_{\rm eff} \approx$ 14,000K.
These results support the hypothesis regarding
the efficiency of atomic diffusion in the stellar atmospheres of
BHB stars with $T_{\rm eff}\succeq$ 11,500~K.
\end{abstract}


\section{Introduction}

According to the current understanding of stellar evolution, the horizontal-branch (HB) stars
are post-main sequence stars that burn helium in their core
and hydrogen in a shell (e.g. Hoyle\& Schwarzschild 1955; Moehler 2004). In this paper we consider the HB stars
that are located in the blue part of the HB, to the left of the RR Lyrae instability strip.
These stars are commonly called blue horizontal-branch (hereafter BHB) stars to distinguish them from
the red horizontal-branch (RHB) stars, which exhibit different observational properties.

\begin{table*}
\centering
\begin{minipage}{150mm}
\caption[]{Journal of Keck+HIRES spectroscopic observations of the
selected hot BHB stars from Behr \shortcite{Behr03a}.} \label{tab1}
\begin{tabular}{@{}lccccccccccc@{}}
\hline
Cluster/Star&     HJD    &Exposure&  S/N & Coverage &$T_{\rm eff}$&$\log{g}$&$V\sin{i}$ &$V_{\rm r}$& [Fe/H] & [He/H]\\
            &2450000+&   Time (s)   & &  (\AA)  &   (K)     &  (dex)    &(km s$^{-1}$)&(km s$^{-1}$)& & \\
\hline
M13/WF2-820 & 1052.7815 &3$\times$1200& 34 & 3888-5356 & 11840 & 3.90 & 4.17 &-239.7$\pm$0.7&-0.22$^{+0.14}_{-0.23}$&-1.09$^{+0.47}_{-0.44}$\\
M13/WF2-2692& 1047.7400 &3$\times$1800& 34 & 3885-6292 & 12530 & 4.08 & 4.64 &-236.2$\pm$0.7&-0.23$^{+0.76}_{-0.34}$&-1.81$^{+0.96}_{-1.07}$\\
M13/WF2-3123& 1248.1298 &3$\times$1200& 35 & 3885-6292 & 13670 & 4.31 & 6.32 &-238.3$\pm$1.1&-0.71$^{+0.65}_{-0.51}$&-1.28$^{+1.55}_{-0.92}$\\
M13/WF3-548 & 1053.7293 &3$\times$1200& 32 & 3888-5356 & 13100 & 4.16 & 4.29 &-235.3$\pm$0.7&-0.64$^{+0.60}_{-0.35}$&-2.10$^{+1.08}_{-0.98}$\\
M13/WF3-1718& 1248.0811 &3$\times$1200& 48 & 3888-5884 & 11510 & 3.79 & 1.89 &-244.0$\pm$0.3&+0.02$^{+0.10}_{-0.13}$&-1.86$^{+0.42}_{-0.68}$\\
M15/B203    & 1408.0180 &3$\times$1200& 28 & 3885-6292 & 13990 & 3.84 & 4.89 & -94.7$\pm$1.2&+0.02$^{+0.38}_{-0.42}$&-2.54$^{+1.15}_{-0.67}$\\
M15/B315    & 1409.0238 &3$\times$1200& 24 & 3885-6292 & 12890 & 3.81 & 1.72 &-103.3$\pm$0.8&+0.04$^{+0.44}_{-0.36}$&-1.75$^{+0.94}_{-0.99}$\\
M15/B334    & 1052.8791 &4$\times$1500& 39 & 3888-5356 & 10750 & 3.61 & 9.22 &-109.3$\pm$1.3&-2.37$^{+0.19}_{-0.21}$&-0.64$^{+0.45}_{-0.45}$\\
M15/B374    & 1406.0183 &3$\times$1200& 21 & 3885-6292 & 12820 & 3.82 & 3.92 &-106.5$\pm$0.9&+0.24$^{+0.48}_{-0.37}$&-1.97$^{+0.94}_{-0.92}$\\
M92/B176    & 1049.0646 &3$\times$1800& 56 & 4020-5520 & 11150 & 3.76 & 6.96 &-116.3$\pm$1.4&-2.25$^{+0.16}_{-0.24}$&-0.55$^{+0.36}_{-0.32}$\\
NGC288/B16  & 1406.0850 &3$\times$1200& 21 & 3885-6292 & 14030 & 4.15 & 3.08 & -44.7$\pm$0.9&+0.66$^{+0.65}_{-0.49}$&-2.82$^{+0.91}_{-0.71}$\\
NGC288/B22  & 1407.0955 &3$\times$1200& 17 & 3885-6292 & 12130 & 4.01 & 2.48 & -43.8$\pm$0.9&+0.28$^{+0.29}_{-0.38}$&-2.36$^{+1.26}_{-0.64}$\\
NGC288/B186 & 1409.0862 &3$\times$1200& 31 & 3885-6292 & 11390 & 3.94 & 3.89 & -43.8$\pm$0.5&+0.17$^{+0.17}_{-0.21}$&-1.45$^{+0.65}_{-0.85}$\\
NGC288/B302 & 1408.0891 &3$\times$1200& 26 & 3885-6292 & 13230 & 4.15 & 1.85 & -43.8$\pm$0.5&+0.61$^{+0.34}_{-0.32}$&-2.63$^{+0.88}_{-0.66}$\\
\hline
\end{tabular}
 \end{minipage}
\end{table*}

BHB stars with $T_{\rm eff} \succeq$ 11,500K show several observational anomalies.
By studying the position of BHB stars of globular
clusters\footnote{Most of the known BHB stars are found in globular clusters.}
on Hertzspruing-Russell diagrams, it is found that they exhibit photometric
jumps (Grundhal et al. 1999) and gaps (Ferraro et al. 1998) at $T_{\rm eff} \simeq $11,500~K. These stars
also show large abundance anomalies (Glaspey et al. 1989; Behr et al. 1999; Moehler et al. 1999; Behr 2003a).
For instance, iron is generally overabundant in these stars relative to their cluster abundance. It is also found
that the spectroscopic gravities of these stars are lower than those predicted by the canonical models
(Crocker et al. 1988; Moehler et al. 1995). It is commonly believed that atomic diffusion of the elements in the
atmosphere of these stars is responsible for the aforementioned observational anomalies.
Atomic diffusion arises from the
competition between radiative acceleration and gravitational settling. This can produce a net acceleration on
atoms and ions, which results in their diffusion \cite{Michaud70} and which may lead to abundance anomalies.

In addition to the detection of vertically stratified abundances in some BHB stars (Khalack et al. 2007 and 2008,
more information is given below), BHB stars with $T_{\rm eff} \succeq$ 11,500K have lower rotational velocities
than their cooler counterparts (Peterson et al. 1995; Behr et al. 2000a and 2000b; Behr 2003b).
These observations suggest that the atmospheres of such stars are stable enough to have atomic diffusion.
Theoretical results of Quievy et al. (2009) show that helium sinks in stars with low rotational velocities, which
leads to the disappearance of the superficial He convection zone. This then opens the door for atomic diffusion
to play a role.

Other theoretical results also concur with the belief that diffusion is present in the atmospheres of BHB
stars with $T_{\rm eff} \succeq$ 11,500K. For instance, the atmospheric models of Hui-Bon-Hoa,
LeBlanc \& Hauschildt \shortcite{Hui-Bon-Hoa+00} showed that the observed photometric jumps and gaps for
hot BHB stars can be explained by elemental diffusion in their atmosphere. These models calculate self-consistently
the structure of the atmosphere while taking into account the stratification predicted by diffusion (assuming equilibrium). These models have recently been improved (LeBlanc et al. 2009) and confirm that vertical stratification of the elements can strongly modify the structure of the atmospheres of BHB stars. Such structural changes for the atmosphere lead to the photometric anomalies discussed above.



As mentioned above, by synthesizing spectral line profiles, Khalack et al. \shortcite{Khalack+07}
found vertical abundance stratification of sulfur in the atmosphere of the field BHB star HD~135485.
Vertical stratification of iron abundance was also detected by Khalack et al. \shortcite{Khalack+08a}
in the atmosphere of
two BHB stars in the globular cluster M15 and for one star in the globular cluster M13. Also, more recently,
Hubrig et al. \shortcite{Hub09} have found isotopic anomalies of calcium in six BHB stars.

In this paper we will attempt to detect signatures of vertical abundance stratification
of iron in a set of fourteen BHB stars from line profile analysis for which we have appropriate
data. This extension of the studies of Khalack et al. (2007 and 2008) will serve to provide a larger amount
of observational data for iron stratification which can then be used to verify theoretical models.
We have selected the BHB stars from the spectra of Behr \shortcite{Behr03a} relying on their value of
effective temperature and their low $V\sin{i}$ ($\leq$ 10 km s$^{-1}$) so that atomic diffusion phenomena
would most probably be present and detectable in these stars.
The properties of the acquired spectra are discussed in Sec.~\ref{obs}, while in Sec.~\ref{mod}
the details concerning the simulation routine and adopted atmospheric parameters
for the program stars are described. The evidence for vertical stratification of iron
is given in Sec.~\ref{vert} along with the estimation of the mean iron abundance, $V\sin{i}$  and
radial velocities for each BHB star considered here. A discussion follows in Sec.~\ref{discuss}.

\section[]{Observations}
\label{obs}

The selected BHB stars were observed with the Keck I telescope and the HIRES spectrograph. The stars were selected
from the large sample of stars observed by Behr \shortcite{Behr03a}. The criteria by which the stars considered
here were chosen is that their $T_{\rm eff}$ is close to or larger than 11,500 K and that they possess
a low $V\sin{i}$ ($\leq$10 km s$^{-1}$).
These criteria were chosen to maximize the chance that atomic diffusion is present in their atmosphere and that
its effects are potentially observable.

Table~\ref{tab1} summarizes the information concerning the observations of the fourteen stars chosen by giving
(in individual columns) the object identification,
the heliocentric Julian Date of the observation, the exposure time, the S/N per pixel
the spectral coverage, the size of the seeing disk (FWHM), the effective temperature,
the logarithm gravity and the heliocentric radial velocity.

The size of the C1 slit is 0.86 arcsec which
provides a spectral resolution of $R=\lambda/\delta\lambda=$45,000.
Behr \shortcite{Behr03a} has found that for the aforementioned stars the underfilling of the slit
should not change the estimated spectral resolution by more than 4\%--7\%.
To process the spectra, Behr~\shortcite{Behr03a} has employed the package of routines developed
by McCarthy \shortcite{McCarthy90} for the FIGARO data analysis package \cite{Shortridge93}.
A comprehensive description of the data acquisition and reduction procedure is presented
by Behr~\shortcite{Behr03a}.

\begin{table}
\begin{minipage}{70mm}
\caption[]{Properties of the BHB stars studied here.} \label{aver}
\begin{tabular}{@{}lcccc@{}}
\hline
Cluster/Star& [Fe/H] & $V\sin{i}$ &$V_{\rm r}$ & $n$\\
            &        &(km s$^{-1}$)&(km s$^{-1}$)& \\
\hline
M13/WF2-820 & -0.09$\pm$0.32 & 4.2$\pm$1.0 &-239.6$\pm$0.8 & 79\\
M13/WF2-2692& +0.03$\pm$0.25 & 4.4$\pm$1.0 &-236.2$\pm$1.0 & 74\\
M13/WF2-3123& -0.57$\pm$0.17 & 6.0$\pm$1.5 &-238.3$\pm$1.0 & 21\\
M13/WF3-548 & -0.49$\pm$0.16 & 4.2$\pm$0.7 &-235.0$\pm$1.3 & 26\\
M13/WF3-1718& +0.14$\pm$0.13 & 2.0$\pm$0.3 &-243.7$\pm$0.7 & 96\\
M15/B203    & -0.14$\pm$0.22 & 5.1$\pm$1.3 & -95.0$\pm$0.3 & 32\\
M15/B315    & +0.10$\pm$0.21 & 1.8$\pm$0.2 &-103.2$\pm$0.9 & 72\\
M15/B334    & -1.38$\pm$0.86 &10.1$\pm$0.5 &-106.5$\pm$2.1 & 11\\
M15/B374    & +0.59$\pm$0.21 & 4.0$\pm$0.9 &-107.0$\pm$1.1 & 82\\
M92/B176    & -1.26$\pm$0.75 & 7.0$\pm$1.0 &-117.7$\pm$1.2 & 21\\
NGC288/B16  & +0.69$\pm$0.15 & 2.9$\pm$0.9 & -45.1$\pm$0.9 & 42\\
NGC288/B22  & +0.52$\pm$0.22 & 2.5$\pm$0.5 & -43.9$\pm$0.8 & 47\\
NGC288/B186 & +0.40$\pm$0.17 & 3.3$\pm$0.8 & -43.9$\pm$0.5 & 65\\
NGC288/B302 & +0.77$\pm$0.17 & 1.8$\pm$0.2 & -44.0$\pm$0.5 & 75\\
\hline
\end{tabular}
 \end{minipage}
\end{table}

\section[]{Simulation procedure}
\label{mod}

The line profile simulations were performed with the {\sc Zeeman2} spectrum
synthesis code (Landstreet \shortcite{Landstreet88}, Wade et al. \shortcite{Wade+01}).
Khalack \& Wade \shortcite{Khalack+Wade06} have modified the {\sc Zeeman2} code to
allow for an automatic minimization of the model parameters using the
{\it downhill simplex method} \cite{press+}.
To analyze the vertical abundance stratification we have built the dependence of the abundance derived from
each analyzed profile relative to the optical depth $\tau_{\rm 5000}$
for the list of selected line profiles, assuming that the profile is formed mainly at
line optical depth $\tau_{\rm \ell}$=1. This depth corresponds to a
continuum optical depth $\tau_{\rm 5000}$ of a certain value that in turn corresponds
to a given layer of the stellar atmosphere model.
A comprehensive description of the procedure is given by Khalack et al.
\shortcite{Khalack+07}. This method was also used by Khalack et al. \shortcite{Khalack+08a} 
for the study of stratification in BHB stars and by Thiam et al. \shortcite{Thiam+10} for HgMn stars.

The stellar atmosphere models used for our analysis were calculated with the {\sc Phoenix} code
\cite{Hauschildt+97} assuming LTE (Local Thermodynamic Equilibrium), solar metallicity
except for the iron abundance and depleted helium abundances, and
the surface gravities 
extracted from Behr \shortcite{Behr03a} which are listed in Table~\ref{tab1}.
The depleted helium abundance has also been taken into account during
line profile simulations using {\sc Zeeman2}.


In a previous study, \cite{Khalack+08a} showed that inclusion of
non-zero microturbulence leads to the amplification of signatures of
vertical abundance stratification for different chemical species.
Therefore, in order to mitigate the influence of microturbulence on
the estimation of iron vertical stratification, we have assumed zero
microturbulent velocity in the simulations presented here.

In this paper, we have selected a list of 
Fe\,{\sc ii} lines that are
suitable for abundance and stratification analysis.
The full list of the analyzed spectral lines is given in the electronic edition of {\it MNRAS}.
Atomic data for Fe\,{\sc ii} lines are extracted from VALD-2
(Kupka~et~al.~\shortcite{Kupka+99}, Ryabchikova~et~al.~\shortcite{Ryab+99}),
Pickering et al. \shortcite{Pickering+01} and 
Raassen \& Uylings\footnote{\rm ftp://ftp.wins.uva.nl/pub/orth} \shortcite{RU+98}) line database.

\section{Iron vertical stratification}
\label{vert}

A sample of fourteen BHB stars from the globular clusters M13, M15, M92 and NGC~288 were
studied using the procedure briefly described in Sec.~\ref{mod}. These stars were chosen using the criteria described
in Sec.~\ref{obs} and some of their properties obtained by our analysis are listed in Table~\ref{aver}, namely,
their average iron abundance, $V\sin{i}$ and mean heliocentric radial velocity averaged over all lines analyzed
($n$ represents the number of lines).
The reported uncertainties are equal to the standard deviation calculated from the results of individual line
simulations for all lines considered. The results shown in Table~\ref{aver} for the iron abundance are consistent
to those found by Behr \shortcite{Behr03a} (see Table~\ref{tab1}) except for the stars B334 and B176. This difference
is partly due to the fact that Behr \shortcite{Behr03a} used mainly strong iron lines for his analysis while we
also use weak lines which are formed deeper in the atmosphere and predict larger average iron abundance. A larger
abundance is obtained because these two stars show an enhancement of iron abundance in the deeper atmospheric layers.
The radial velocities found here are also consistent for all the stars, but once again the stars B334 and B176 show a
small difference between our results and those of Behr \shortcite{Behr03a}. These two stars will be discussed in more
detail below.

The effective temperatures of the BHB stars studied here
range from 10,750K to 14,030K (see Tab.~\ref{tab1}). This and the variety of the
mean abundance values for iron (see Tab.~\ref{aver}) among the sample of BHB stars
provides a unique opportunity to search for evidence of abundance vertical stratification of iron and to
estimate its possible relation to other physical characteristics of these stars.

\begin{table}
\centering
\begin{minipage}{70mm}
\caption[]{Data on vertical stratification of iron abundance.}
\label{tab3}
\begin{tabular}{@{}l|cc|ccc@{}}
\hline
Cluster/Star& \multicolumn{2}{c}{FeII} & $n$\\
            &slope&$\log(\frac{N_{Fe}}{N_{tot}})$& \\
\hline
M13/WF2-820 & 0.30$\pm$0.02&-4.59$\pm$0.32 & 79\\
M13/WF2-2692& 0.20$\pm$0.02&-4.47$\pm$0.25 & 74\\
M13/WF2-3123& 0.05$\pm$0.07&-5.07$\pm$0.17 & 21\\
M13/WF3-548 & 0.04$\pm$0.04&-4.99$\pm$0.16 & 26\\
M13/WF3-1718& 0.02$\pm$0.02&-4.36$\pm$0.13 & 96\\
M15/B203    & 0.15$\pm$0.04&-4.64$\pm$0.22 & 32\\
M15/B315    & 0.07$\pm$0.03&-4.40$\pm$0.21 & 72\\
M15/B334    & 0.72$\pm$0.08&-5.89$\pm$0.87 & 11\\
M15/B374    & 0.07$\pm$0.04&-3.91$\pm$0.21 & 82\\
M92/B176    & 0.65$\pm$0.07&-5.76$\pm$0.75 & 21\\
NGC288/B16  & 0.01$\pm$0.03&-3.81$\pm$0.15 & 42\\
NGC288/B22  & 0.12$\pm$0.04&-3.98$\pm$0.22 & 47\\
NGC288/B186 & 0.13$\pm$0.02&-4.10$\pm$0.17 & 65\\
NGC288/B302 & 0.06$\pm$0.02&-3.73$\pm$0.17 & 75\\
\hline
\end{tabular}
 \end{minipage}
\end{table}

\subsection{Search for signatures of vertical stratification}
\label{trac}

\begin{figure*}
\includegraphics[scale=0.31,angle=-90]{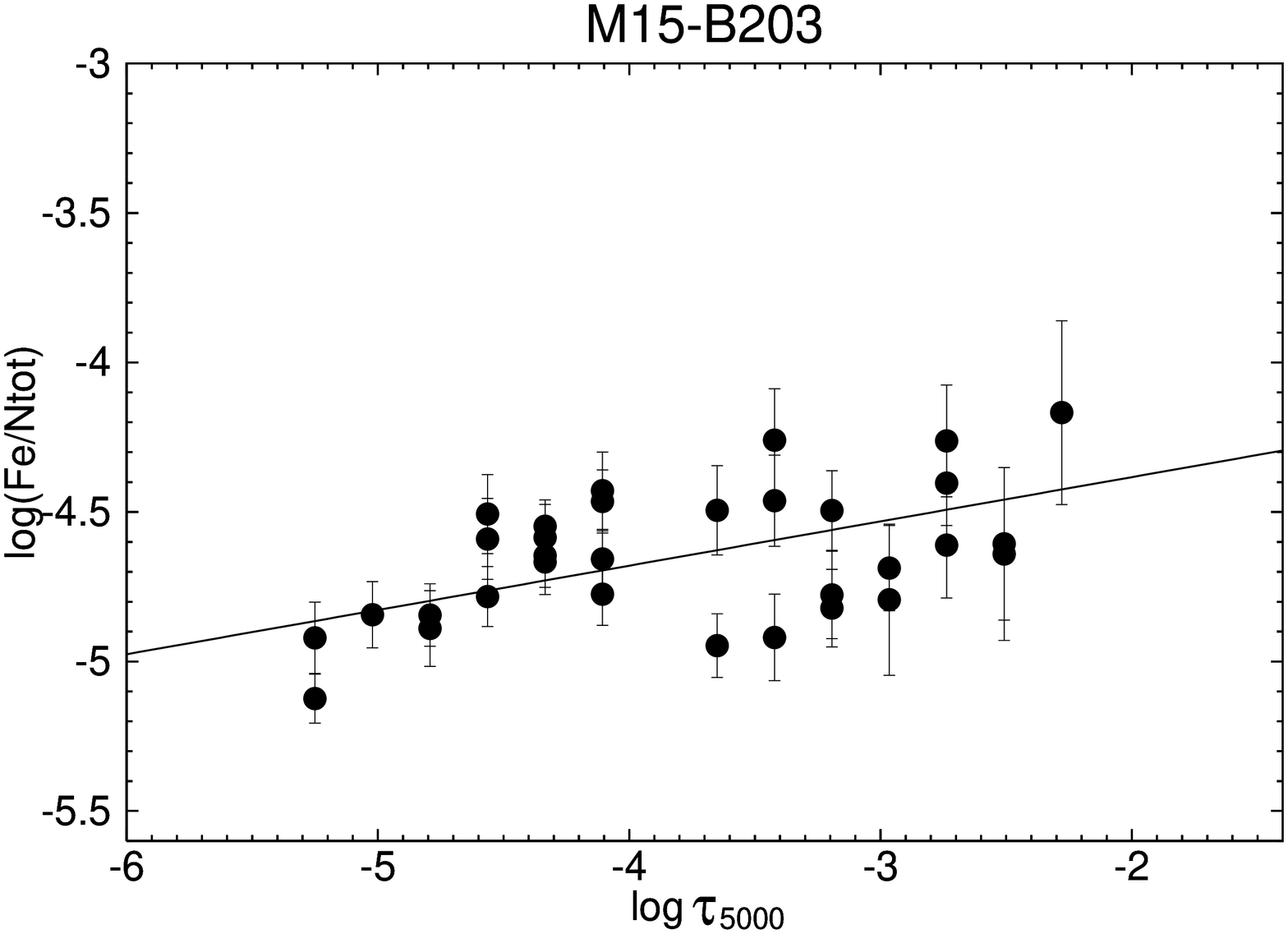}\\
\includegraphics[scale=0.31,angle=-90]{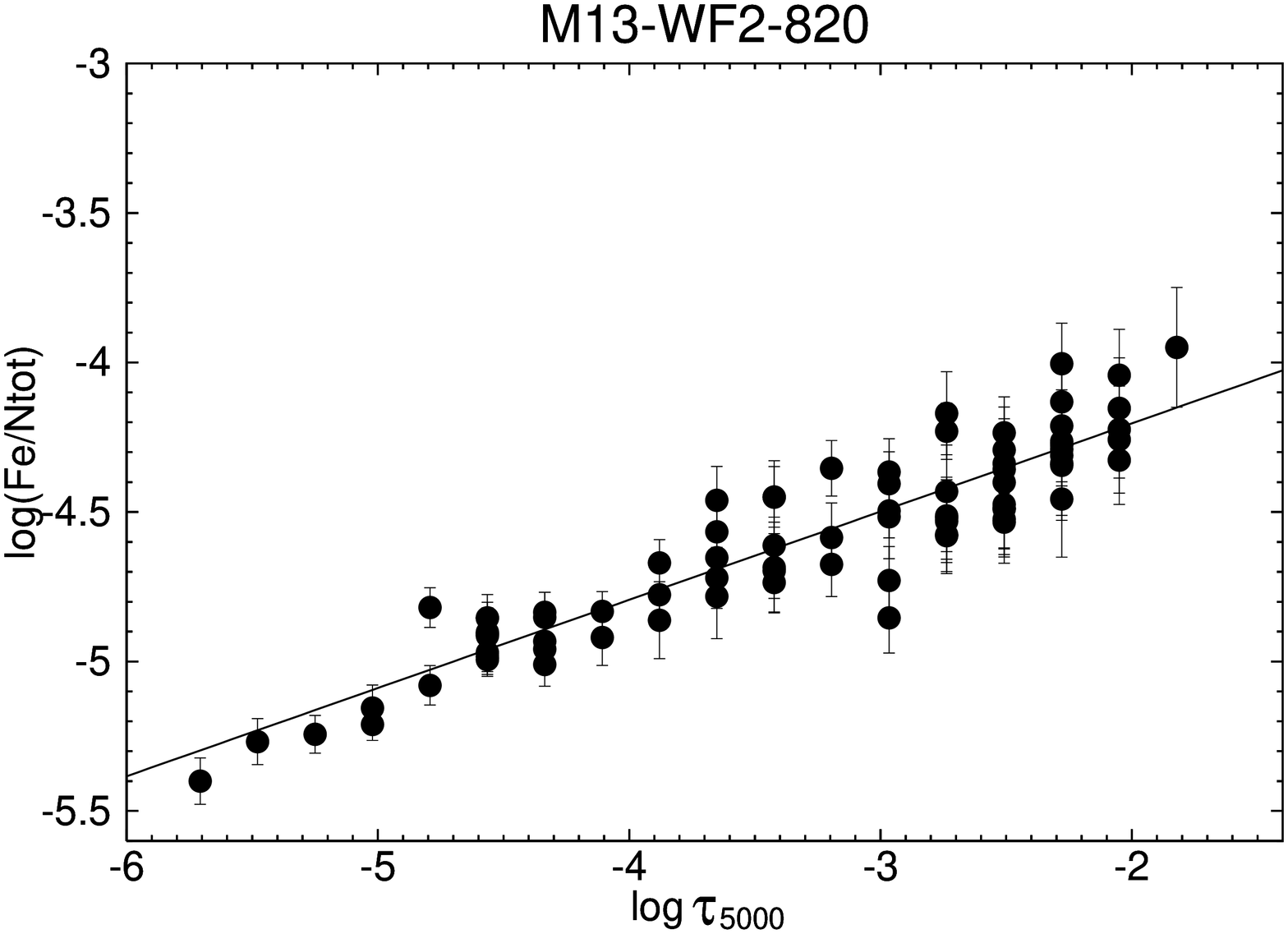}
\includegraphics[scale=0.31,angle=-90]{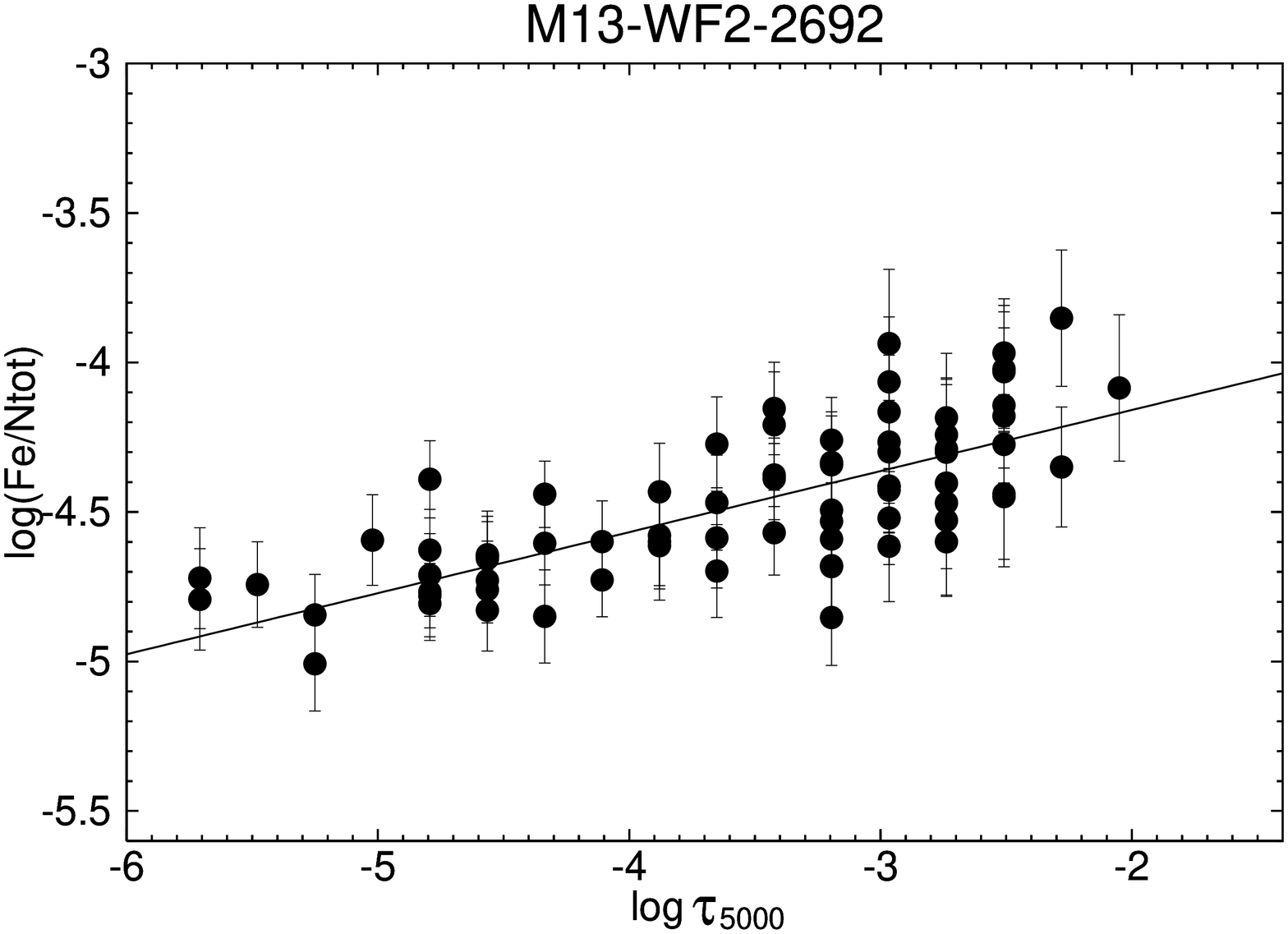}\\
\includegraphics[scale=0.31,angle=-90]{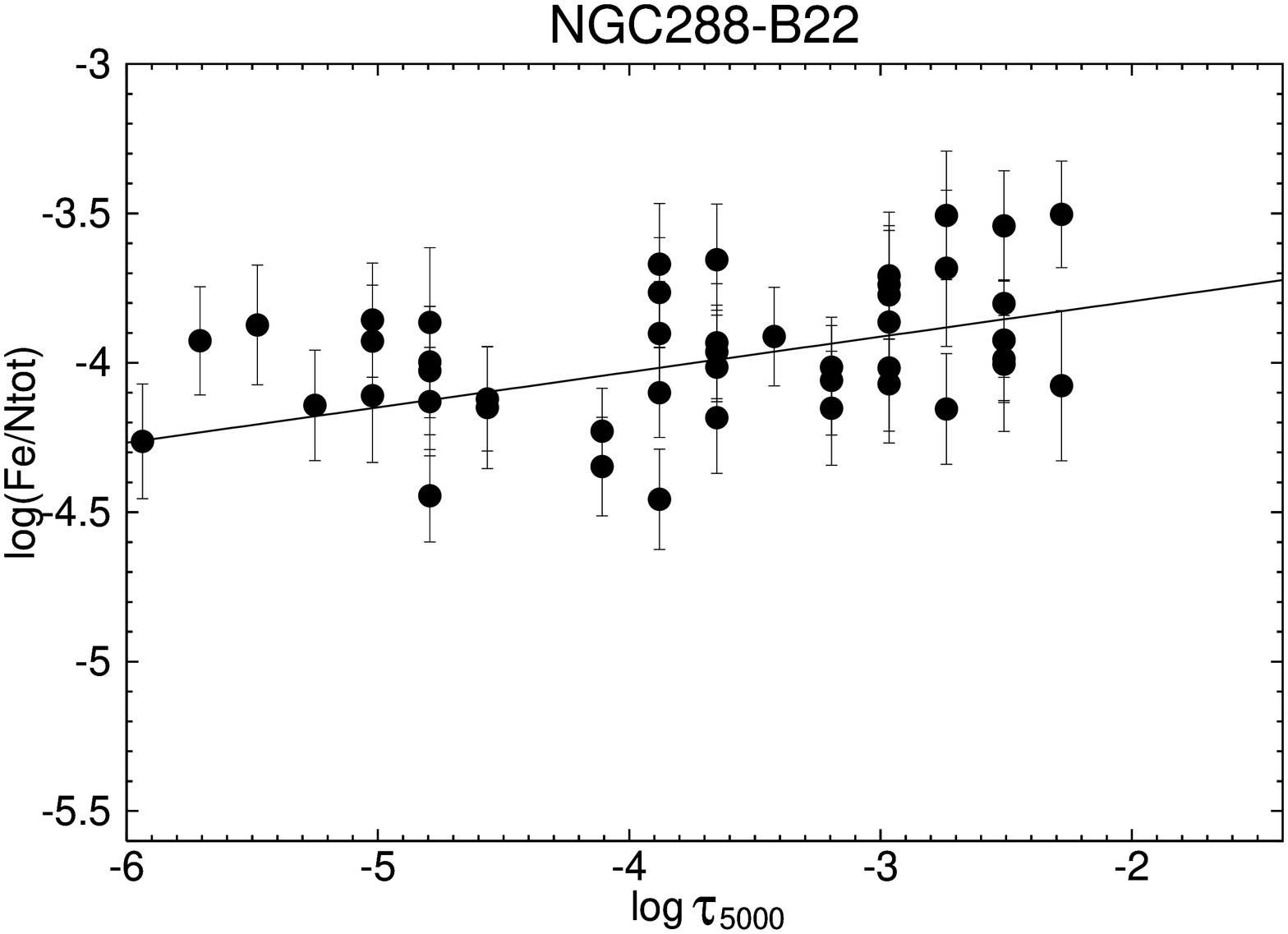}
\includegraphics[scale=0.31,angle=-90]{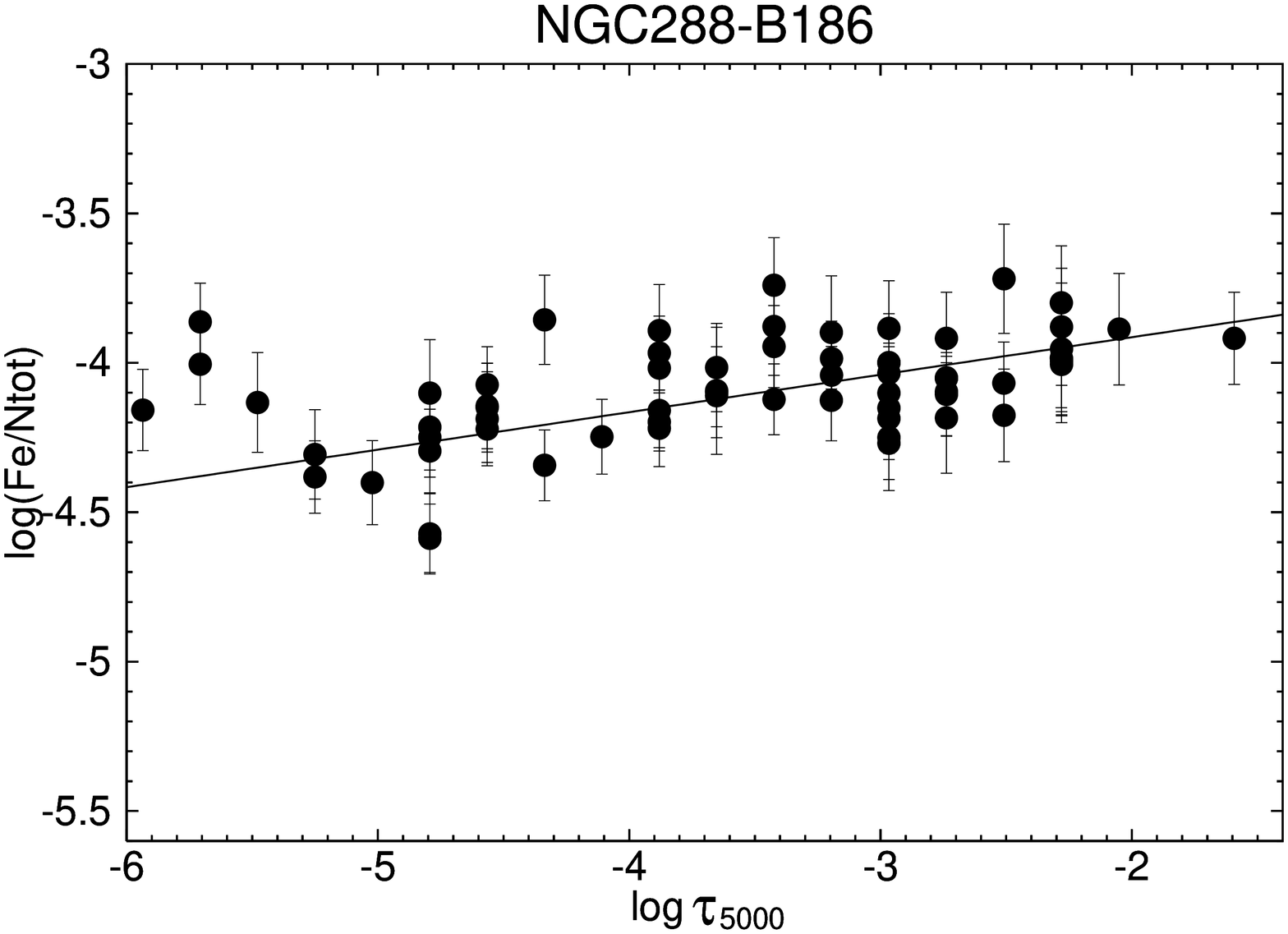}\\
\includegraphics[scale=0.31,angle=-90]{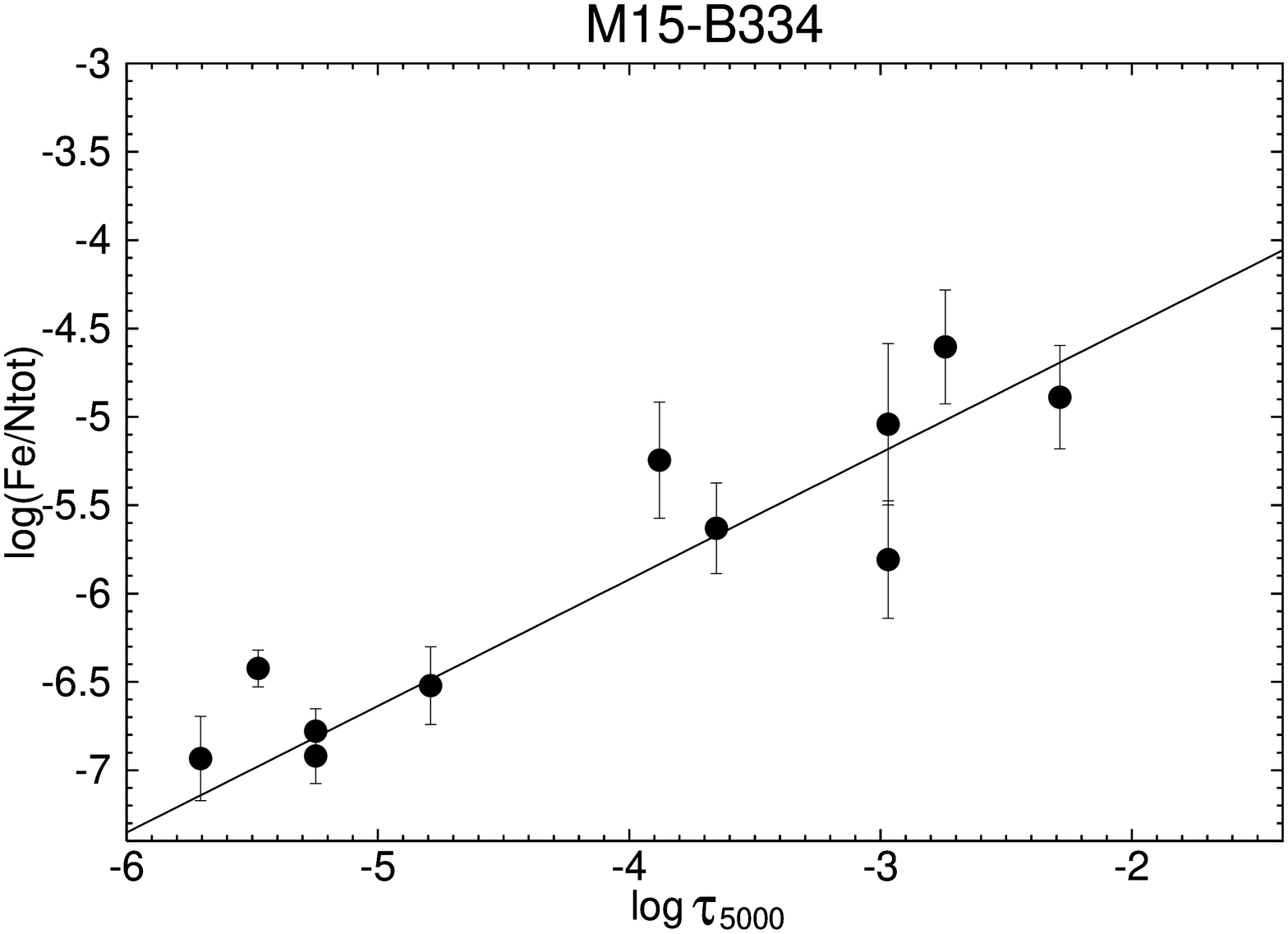}
\includegraphics[scale=0.31,angle=-90]{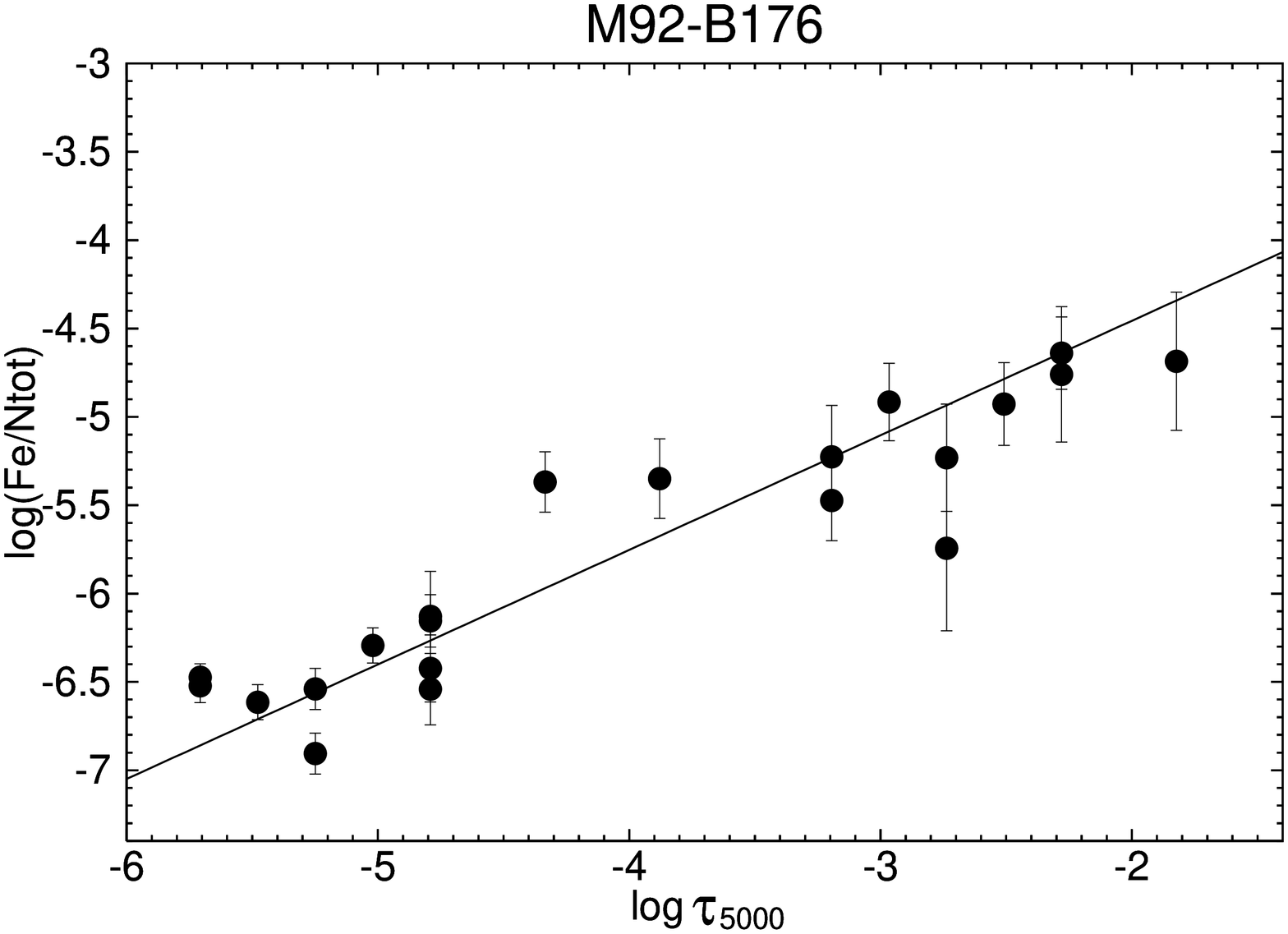}
\caption{ Abundance estimates from the analysis of Fe\,{\sc ii}
(filled circles) lines as a function
of line (core) formation optical depth 
for BHB stars B203 in M15 (middle top), WF2-820 (second row left) and
WF2-2692 (second row right) in M13, B22 (third row left) and B186 in NGC 288
(third row right), B334 in M15 (bottom left) and B176 in M92 (bottom right).
The solid line approximates the data with a
linear fit using the least-square algorithm.} \label{fig1}
\end{figure*}

To verify if iron abundance is vertically stratified in the stellar atmospheres of the selected BHB stars,
we have analyzed individually all Fe\,{\sc ii} line profiles present in the spectra that give
small errors
for the estimated parameters (iron abundance, $V_{\rm r}$ and $V\sin{i}$), assuming zero microturbulence. The
measured abundance for each line was then associated to a line depth formation.
The iron abundances obtained for different optical depths (see Figs.~\ref{fig1}) were fitted to a linear function
(for the range of atmospheric depths $-5.3 < \log{\tau_{5000}} < -2.0$) using the least-square algorithm to statistically
evaluate the significance of observable trends. This range of $\log{\tau_{5000}}$ was chosen because spectra from our
sample of stars all possess iron lines within this interval. The main result found here is that the iron abundance generally increases
towards the deeper atmosphere, although some of the increases found are not statistically significant. The slope of
these gradients are presented in Table~\ref{tab3}, where the individual columns give
the increase of $\log{N_{Fe}/N_{tot}}$ (in dex) calculated per dex of $\log{\tau_{5000}}$, the average
iron abundance value obtained taking into account all of the lines and the number of lines analyzed.

Without further analysis, it is found that among our sample of fourteen stars, seven stars show clear signatures
of vertical stratification of iron. Two of them (B22, B186) are located in the globular cluster NGC~288,
two (WF2-820, WF2-2692) in M13, two (B203, B334) in M15 and one (B176) in M92 (see Fig.~\ref{fig1} and Tab.~\ref{tab3}).
One more BHB star, B302 (in NGC~288) also shows a slope that is statistically significant
(with respect to the error bar), but its value is too low to confidently state, due to all of the uncertainties
inherent to line analysis, that iron stratification exists in this star.

However, for two of the seven BHB stars that show iron stratification (B334, B176), iron is strongly depleted
(in comparison with the solar abundance of iron) and the number of analyzed iron lines is therefore respectively
smaller than for most of the other stars studied here (see Tab.~\ref{tab3}).
However, the iron abundance at $\log{\tau_{5000}}$ = -2 is nearly solar and is therefore higher than the cluster abundance.
The same two stars show very large slopes for the iron abundance with respect to optical depth (see bottom row of Fig.~\ref{fig1}).
Since our analysis is based on a relatively small number of lines (especially for B334), these results could
be distorted by few cases of line misidentification. It is also surprising that such a large stratification
slope is found where iron abundance is not enhanced in the outer atmosphere and helium abundance is large.
If diffusion is at play, a larger iron abundance and a smaller
helium abundance would be expected.
%
%
These two stars also appear to have radial velocities, determined from the analysis of Fe\,{\sc ii} lines,
slightly different (even though this difference is inside the error bars it is greater than for the other stars of
our sample) than those obtained by Behr \shortcite{Behr03a} from
the combined analysis of lines of different chemical species. The average iron abundance found here for these two
stars (see Tab.~\ref{aver}) are quite different form those found by Behr \shortcite{Behr03a} (see Tab.~\ref{tab1}),
but this is due to the fact that we have also studied some weaker lines (see Sec.~\ref{trac}).
Therefore, all of these discrepancies cast some doubts about our results for these two stars and therefore more
high-quality spectra is needed to confirm the presence of iron stratification in their atmosphere.

The vertical iron stratification for the other five stars:
WF2-820 and WF2-2692 in M13, B22 and B186 in NGC~288 and B203 in M13 is shown in Fig.~\ref{fig1}.
The lines used to produce these
figures span several orders of magnitude for $\tau_{5000}$. Also, since the increase of the abundance along this
depth range is too large to be explained by the various uncertainties intervening in abundance determination,
it is safe to state that iron is vertically stratified in these stars.

\subsection{Dependence of stratification on effective temperature}
\label{Teff}

It is clear from the results shown in the previous section and those of Khalack et al. (2007 and 2008),
that some BHB stars show signatures of vertical stratification of iron abundance, while others
do not reveal such a property.
All the studied stars (except B334, which has $T_{\rm eff}\simeq$10,750K) have $T_{\rm eff}$ higher than 11,000K
and are therefore in or near the
domain of $T_{\rm eff}$ where diffusion is expected to come into play. However, these stars are found in different
globular clusters and are shown to have different mean iron abundances.
The effective temperature is one of the most important parameter for a stellar atmosphere and has
direct impact on the atomic diffusion in BHB stars (Hui-Bon-Hoa et al. 2000 and LeBlanc et al. 2009).
It would therefore be very interesting to verify if a dependence of vertical iron stratification on $T_{\rm eff}$
exists in BHB stars.

To track possible dependence of iron vertical stratification on $T_{\rm eff}$ we have combined the results obtained
in this study with the previously published results
by Khalack et al.~\shortcite{Khalack+08a} for other BHB stars,
whose iron abundance slope (see below) were recalculated for the same range of atmospheric
depths $-5.3 < \log{\tau_{5000}} < -2.0$.

The slope of the iron abundance as a function of optical depth for the various BHB stars was calculated and
a correlation is found between this slope and $T_{\rm eff}$. Fig.~\ref{fig4} shows the iron slopes
with respect to $T_{\rm eff}$, along with a linear fit for these data-points using the least-square
algorithm (solid line). The stars B334 in M15 and B176 in M92 were excluded from this fit because of the
reasons discussed in Subsec.~\ref{trac}. They are however shown in Fig.~\ref{fig4} as
empty circles. The inclusion of these two stars would not change the qualitative conclusion that the slope of iron
abundance decreases as a function of $T_{\rm eff}$ for BHB stars with $T_{\rm eff}\succeq$ 11,500~K.
The BHB star WF3-1718 in M13 (shown as empty square in Fig.~\ref{fig4}) was also not taken into account during
the aforementioned fit depicted by the solid line, because the wings of
Balmer lines in its spectrum suggest that a higher $T_{\rm eff}$ is probable for this star. Nevertheless,
inclusion of WF3-1718 still results in some visible correlation (see dashed line in Fig.~\ref{fig4}). Adoption of
higher effective temperature for WF3-1718 would amplify this correlation.


The detected correlation indicates that the strongest vertical stratification of iron abundance is expected
for BHB stars with $T_{\rm eff}$ near 11,500K while it decreases for higher values of $T_{\rm eff}$.
It becomes negligibly small for the BHB stars with $T_{\rm eff} \succeq$ 14,000K.
LeBlanc et al. (2010, in preparation) find that the theoretical models for BHB stars of LeBlanc et al. (2009)
predict a similar tendency of decreasing iron vertical abundance with $T_{\rm eff}$.



\section{Discussion}
\label{discuss}

\begin{figure}
\includegraphics[scale=0.32,angle=-90]{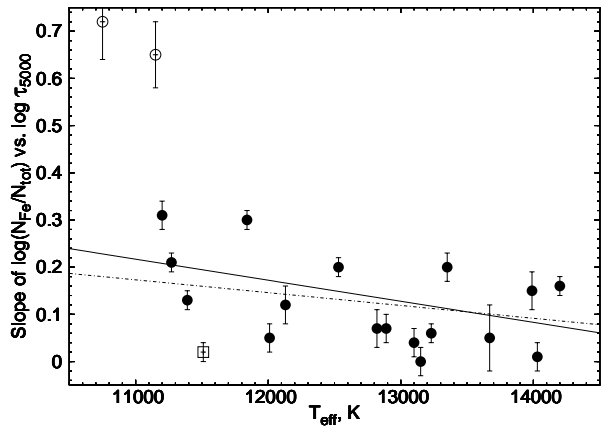}
\caption{Plot of the slope of iron vertical stratification versus effective temperature of analyzed BHB stars
(see text for details). The dashed line is a linear fit that includes all stars except for the two stars represented
by empty circles (B176 and B334). The linear fit given by the solid line also excludes the star WF3-1718
represented by the empty square.}
\label{fig4}
\end{figure}

In this paper, we have found additional evidence that iron abundance is vertically stratified in
the atmosphere of BHB stars. From the analysis of 14 BHB stars, selected as the most probable candidates for such
detections
due to their $T_{\rm eff} \succeq$ 11,500K and $V\sin{i} \leq$ 10 km s$^{-1}$, we have found clear signatures of
iron stratification in five stars: B22, B186 in the globular cluster NGC~288; WF2-820, WF2-2692 in M13 and
B203 in M15. Two other stars, B334 in M15 and B176 in M92, show extremely strong increase of iron abundance
as a function of optical depth in their atmosphere (see Tab.~\ref{tab3} and Fig.~\ref{fig4}). However, this result
may be influenced by the strongly depleted mean iron abundance (that results in a smaller number of analyzed
line profiles). Also, because of other anomalies found for these two stars (see Sect.~\ref{trac}), some doubts
remain about observed stratification of iron in their atmosphere.

For B334 and B176 the effective temperature is only slightly smaller than the limit ($T_{\rm eff}=$11,500~K)
above which BHB stars show the photometric jumps and gaps (Grundahl et al.~\shortcite{Grundahl+99},
Ferraro et al.~\shortcite{Ferraro+98}) and where diffusion is efficient.
On the other hand, Khalack et al.~\shortcite{Khalack+08a} have shown that stellar atmosphere models with higher
$T_{\rm eff}$ usually slightly decrease the slope of the element's abundance with respect to the logarithm
of excitation potential as well as with respect to $\log{\tau_{5000}}$. Therefore, adoption of higher
effective temperatures for B334 and B176 will decrease the estimated slopes of iron abundance which increases towards
the deeper atmosphere.

The values of effective temperature for the sample of analyzed BHB stars were derived from
Kurucz's ATLAS9 synthetic photometry grids \cite{Behr03a}, which use the stellar atmosphere models with
homogeneous distribution of element's abundance. Hui-Bon-Hoa et al.~\shortcite{Hui-Bon-Hoa+00} and
LeBlanc et al. (2010, in preparation) have shown that the observable photometric jumps and gaps can be explained
in terms of atomic diffusion mechanism that
leads to stratification of element's abundance with atmospheric depths. The models with stratified abundance
of chemical species result in different photometric colors as compared to chemically homogenous models.
This fact may introduce some errors in the evaluation of $T_{\rm eff}$ by photometric means when using chemically
homogeneous atmospheric models. The uncertainty on $T_{\rm eff}$ determination should be studied by using the new
atmospheric models of LeBlanc et al. (2009) that take into account vertical stratification of the elements.
This is outside the scope of the present paper.

Our analysis of the dependency of the available data for vertical iron stratification in BHB stars
with respect to the effective temperature results in the detection of a trend (see Fig.~\ref{fig4}). The same
results were obtained from the modeling of iron vertical stratification employing the
self-consistent stellar atmosphere models with vertically stratified abundance of chemical species
(LeBlanc et al. 2010, in preparation). The results presented here show that for BHB stars with $T_{\rm eff}$ around 11,500K have
the strongest vertical iron stratification gradients. This gradient diminishes as $T_{\rm eff}$ increases up
to $T_{\rm eff} \simeq$ 14,000K, where no significant iron stratification exists. Nevertheless,
to statistically improve the registered trend of iron stratification more data for BHB stars with appropriate values of
$T_{\rm eff}$ and $V\sin{i}$, especially for $T_{\rm eff}$ near 11,500K need to be analyzed. Such additional data
will be useful to constrain theoretical model atmospheres that include stratification of elements, similar to those
of LeBlanc et al. (2009).


\section*{Acknowledgments}

This research was partially funded by the Natural Sciences and Engineering Research Council
of Canada (NSERC). We thank the R\'eseau qu\'eb\'ecois
de calcul de haute performance (RQCHP) for computing resources.
BBB thanks all the dedicated people involved in the construction and
operation of the Keck telescopes and HIRES spectrograph. He is also
grateful to Judy Cohen, Jim McCarthy, George Djorgovski, and Pat C\^ot\'e
for their contributions of Keck observing time.



\label{lastpage}

\end{document}